# Characterization of elastic topological states using dynamic mode decomposition

Shuaifeng Li[1], Panayotis G. Kevrekidis[2], Jinkyu Yang[1]

[1]Department of Aeronautics and Astronautics, University of Washington, Seattle, WA, USA

[2]Department of Mathematics and Statistics, University of Massachusetts, Amherst, MA, USA

**Abstract**

Elastic topological states have been receiving increased intention in numerous scientific and engineering fields due to their defect-immune nature, resulting in applications of vibration control and information processing. Here, we present the data-driven discovery of elastic topological states using dynamic mode decomposition (DMD). The DMD spectrum and DMD modes are retrieved from the propagation of the relevant states along the topological boundary, where their nature is learned by DMD. Applications such as classification and prediction can be achieved by the underlying characteristics from DMD. We demonstrate the classification between topological and traditional metamaterials using DMD modes. Moreover, the model enabled by the DMD modes realizes the prediction of topological state propagation along the given interface. Our approach to characterizing topological states using DMD can pave the way towards data-driven discovery of topological phenomena in material physics and more broadly lattice systems.

# I. INTRODUCTION

Wave propagation is a typical spatiotemporal phenomenon, which is ubiquitous across science and engineering, especially in fluid dynamics [1,2], geoscience [3,4], plasmas [5], optics [6], atomic and condensed matter physics [7], as well as the more recent field of metamaterials [8–10]. Topological metamaterials have attracted considerable attention not only because of their theoretical significance but also for practical purposes related to materials applications. Wave propagation in elastic topological metamaterials has prominent applications, such as information transmission and vibration control, due to the topological protection [11–16].

The computations involving the propagation of associated waveforms rely mostly on numerical discretization, e.g., finite element and discrete element methods, rather than analytic closed-form solutions which are rarely available in exact form. This naturally generates high-dimensional representations of the solution to accurately reflect the underlying dynamics in both time and space [11,13,14,17]. However, this may occasionally be in contrast with the low-dimensional nature of the underlying dynamics and poses a computational challenge, especially in higher-dimensional settings. Thus, while theoretical modeling and numerical calculation of elastic topological states have been extensively reported, their data-driven analysis and modeling remains far less explored. It is the purpose of the present work to offer a step forward towards filling that void.

Reduced-order models offer representations of the spatiotemporal wave propagation based on the inherently low-rank structure of the simulation data. Within the palette of relevant techniques, dynamic mode decomposition is a powerful dimensionality reduction method to create reduced-order models which identifies spatiotemporal coherent structures from high-dimensional data [18]. Besides, it offers a modal decomposition, where each mode contains spatially correlated structures with the same linear behavior in time, such as oscillations at a certain frequency with growth or decay. Compared with one of the most commonly used dimensionality reduction methods, proper orthogonal decomposition, DMD demonstrates not only dimensionality reduction, but also a reduced model that accounts for how these modes evolve over time. Lately, DMD has been successfully applied to fluid dynamics [18,19], control [20], robotics [21] and neuroscience [22,23]. Hence, developing such an approach for wave propagation in topological metamaterials is highly desirable.

Here, we develop a data-driven framework using DMD for identifying interpretable low-dimensional representations for wave propagation in elastic topological metamaterials created by a select mass-spring system example. The low-dimensional spatiotemporal coherent structures of topological state propagation in our system are extracted, among which correspond to the topological edge states inside the bandgap region. These spatiotemporal coherent structures allow for the reconstruction of the topological state propagation. Moreover, we first demonstrate how to classify the topological and traditional metamaterials using DMD modes via unsupervised clustering. Furthermore, a portion of the data, referred to as the training data, is used to predict the future evolution of the topological states of interest along an interface with arbitrary shape. Our study provides a computationally tractable data-driven characterization of the relevant states and their propagation, paving the way towards the classification and prediction of wave propagation in elastic metamaterials.

Our presentation hereafter will be structured as follows. In section II we will provide a concise introduction in the mathematical and computational details of the DMD algorithm (including technical modifications to the standard algorithm such as the use of a stacking data matrix leveraged herein) and illustrate how it can be used to represent the wave dynamics. In section III, we use DMD to distinguish between the classification of the metamaterials as topological or traditional, while in section IV, we use the algorithm in order to be able to predict future propagation along a topological interface. Finally, in section V, we summarize our findings and provide some direction for future study. The appendices offer details about the band structure of the physical system under consideration, and about further technical aspects of the DMD implementation, such as the DMD spectra and the application of time-delay embedding.

## II. CHARACTERIZATION BY DYNAMIC MODE DECOMPOSITION

To demonstrate DMD on the wave propagation in the elastic topological metamaterial system of interest, we first construct the topological valley metamaterials using spring-mass system, which is realized by alternating the masses at different sites of the unit cell of a mechanical graphene-like lattice [14,17,24,25]. The unit cell contains two different masses $m_1$, $m_2$ and linear spring constant $k_{spring}$. The unit cell has four degrees of freedom specified by the displacement of $m_1$ and $m_1$ ($U = [u_x^{m_1}, u_y^{m_1}, u_x^{m_2}, u_y^{m_2}]$). After periodic boundary conditions are applied on the unit cell, the band structure of our elastic metamaterial can be calculated by identifying the dispersion relation for wave vectors $k$ within the first Brillouin zone:

$$[D(k) + \omega^2 M]U = 0 \qquad (1)$$

Here, $\omega$ denotes the angular frequency of the propagating wave. $M$ is the mass matrix and $D(k)$ is the stiffness matrix as a function of Bloch wave vector $k$. The details are shown in APPENDIX A.

As shown in FIG. 1(a), our system bears a Z-shaped interface which is formed by combining metamaterials with two opposite topological phases. In FIG. 1(a), one such interface is formed by $m_1 = 0.8$ kg (blue), $m_2 = 1.2$ kg (red) shown in the bottom, while the other is formed by $m_1 = 1.2$ kg (red), $m_2 = 0.8$ kg (blue) shown in the top, i.e., with the masses flipped. The spring constant is fixed to be $10^5$ N/m. The schematic of our elastic metamaterial and band inversion process are shown in Appendix A [see, specifically, FIG. 6(a) and FIG. 6(b)]; The projected band structure along the wave vector is also illustrated in the Appendix A [see FIG. 6(c)], where topological states with two types of pseudospins can be observed inside the bandgap.

The system is excited by an oscillating force with the angular frequency of 400 rad/s at the input port of the Z shape interface shown in FIG. 1(a). The masses at the boundaries of the system are connected to springs fixed to the wall, i.e., $\vec{F}_{boundary} = -k_{spring}\vec{u}$, where $\vec{u}$ contains horizontal displacement $u_x$ and vertical displacement $u_y$. Because of the topological protection of wave propagation, the elastic wave can travel through the sharp bend robustly, the horizontal displacement $u_x$ of which is visualized as the time-series snapshots in FIG. 1(b). We then construct the data matrix for DMD by stacking the horizontal displacement $u_x$ and vertical displacement $u_y$, resulting in a $2n \times m$ matrix shown in FIG. 1(b), where $n$ is the number of masses and $m$ is the number of used snapshots over time ($n = 2700$ and $m = 379$ for Z-shape interface based on the numerical simulations of 758 ms duration with 2 ms time interval).

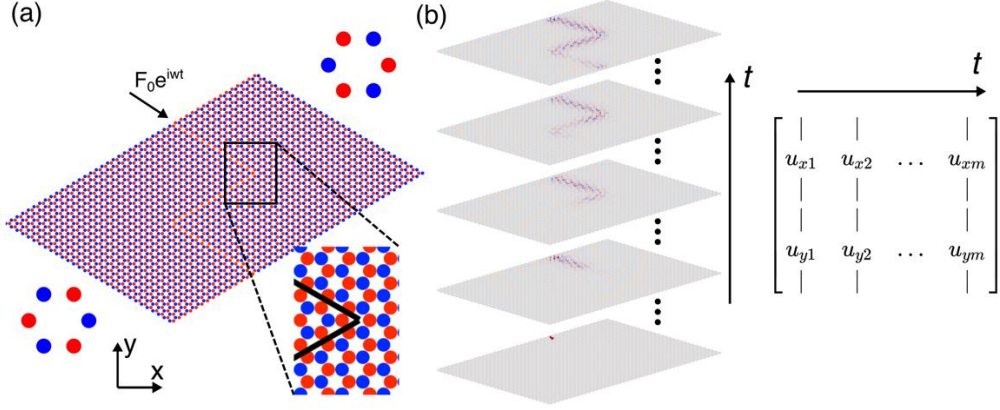

FIG. 1. Setup for simulation and for the numerical DMD implementation. (a) Simulation of wave propagation along the Z-shaped interface in a valley topological metamaterial built by means of a spring-mass system. Two different unit cells with two different topological phases are shown on the two sides. A magnified view of the topological boundary (black line) is shown in the inset. (b) Snapshots of wave propagation represented by the horizontal displacement $u_x$ along Z-shape interface at $t = 2$ ms, 192 ms, 380 ms, 570 ms and 758 ms (with the time evolving from the bottom to the top). The data matrix for DMD is organized by stacking horizontal displacement $u_x$ and vertical displacement $u_y$.

After forming the data matrix, we can construct two $2n \times (m-1)$ submatrices:

$$X = \begin{bmatrix} | & | & & | \\ x_1 & x_2 & \cdots & x_{m-1} \\ | & | & & | \end{bmatrix} \quad X' = \begin{bmatrix} | & | & & | \\ x_2 & x_3 & \cdots & x_m \\ | & | & & | \end{bmatrix} \tag{2}$$

where $\begin{bmatrix} | \\ x_m \\ | \end{bmatrix} = \begin{bmatrix} | \\ u_{xm} \\ | \\ | \\ u_{ym} \\ | \end{bmatrix}$ for simplicity of description. $X$ and $X'$ may be related by a best-fit linear

operator $A$ that minimizes the Frobenius norm error $\|X' - AX\|_F$ given by:

$$X' = AX \implies A = X'X^\dagger \tag{3}$$

where $X^\dagger$ is the Moore-Penrose pseudo-inverse [26]. Because $2n \gg m$ for our systems, so, instead of obtaining $A$ directly, we seek for the eigen decomposition of $A$. After $X$ is decomposed using singular value decomposition (SVD) and the proper rank-$r$ truncation is chosen so that $\tilde{X} = \tilde{U}\tilde{\Sigma}\tilde{V}^T$, where $\tilde{U} \in \mathbb{R}^{2n \times r}$, $\tilde{\Sigma} \in \mathbb{R}^{r \times r}$ and $\tilde{V} \in \mathbb{R}^{(m-1) \times r}$ are the left unitary matrix, diagonal matrix with singular values, and right unitary matrix, respectively, the matrix representation $\tilde{A}$ can be written as:

$$\tilde{A} = \tilde{U}^* X' \tilde{V} \tilde{\Sigma}^{-1} \tag{4}$$

where the $*$ denotes the conjugate transpose. The eigen-decomposition of $\tilde{A}$ results in the matrix of eigenvectors $W$ and eigenvalues $\lambda$, which are the DMD eigenvalues. This further derives the corresponding DMD mode $\phi$, which is the column of $\Phi = X'\tilde{V}\tilde{\Sigma}^{-1}W$.

As discussed in Appendix B, the rank-$r$ truncation is chosen to be $r = 131$ to minimize the reconstruction error and also to eliminate the noise in the simulation data. Each DMD mode $\phi$ corresponds to eigenvalue $\lambda$. The temporal dynamics, referring to growth/decay and the frequency of oscillation of each DMD mode $\phi$, is reflected through the magnitude and phase of eigenvalue $\lambda$, respectively. In our case, because the raw data is strictly real valued, the decomposition yields complex conjugate pairs of eigenvalues and modes.

In FIG. 2(a), the eigenvalues $\lambda$ are visualized on the unit circle in the complex plane, suggesting the corresponding modes are oscillating with certain frequencies. The frequencies are defined as $\omega = \left|\text{imag}(\frac{\log \lambda}{\Delta t})\right|$ and the mode amplitudes are defined as $P = ||\phi||_2^2$, which is the squared $L_2$-norm of the DMD modes. FIG. 2(b) gives the DMD spectrum which provides specific spatial modes in our system for different frequencies. It is obvious that there is a region with large mode amplitudes corresponding to the bandgap region (shaded area). The mode with the largest amplitude inside the bandgap region is chosen as the prototypical mode used to visualize the motion of our system. The horizontal displacement $u_x$ is chosen for the description below.

FIG. 2(c) exhibits the magnitude of this most dominant DMD mode of our system, also showcasing the interface-involving dynamics. The decaying magnitude from the input along the Z-shape interface is due to the constant force excitation at the input. The characteristics of interface states present concentrated displacement along the Z-shape interface and rapid decay away from the interface. Besides, the DMD modes showcase the fact that the elastic wave can travel along an interface featuring bends. Apart from the magnitude, the phases of the DMD mode also reflect the important characteristics of our topological valley metamaterials, as shown in FIG. 2(d). The distribution of phase along the interface has a certain alignment, where the magnified view around the interface clearly shows the phase difference around the corners of the honeycomb, indicating specific valley polarization along a certain topological interface. The counterclockwise and clockwise phase evolutions are shown on three corners and the other three corners of the honeycomb, suggesting the valley pseudospin of the excited valley in our system. The valley pseudospin here refers to the phase difference of DMD modes around the corners of honeycomb. Note that the other DMD modes inside the bandgap have a similar pattern and it is these modes that will be primarily used to reconstruct the dynamical evolution below.

The DMD spectra of wave propagation in topological metamaterials with a straight interface and a cross-shaped interface will be further illustrated in Appendix C. Similarities between the DMD spectra and modes can be found in topological state propagation along different interfaces including the high amplitude inside the bandgap and the topological interface states reflected by the DMD modes. This demonstrates the ability of the DMD to robustly discover the nature of topological state propagation.

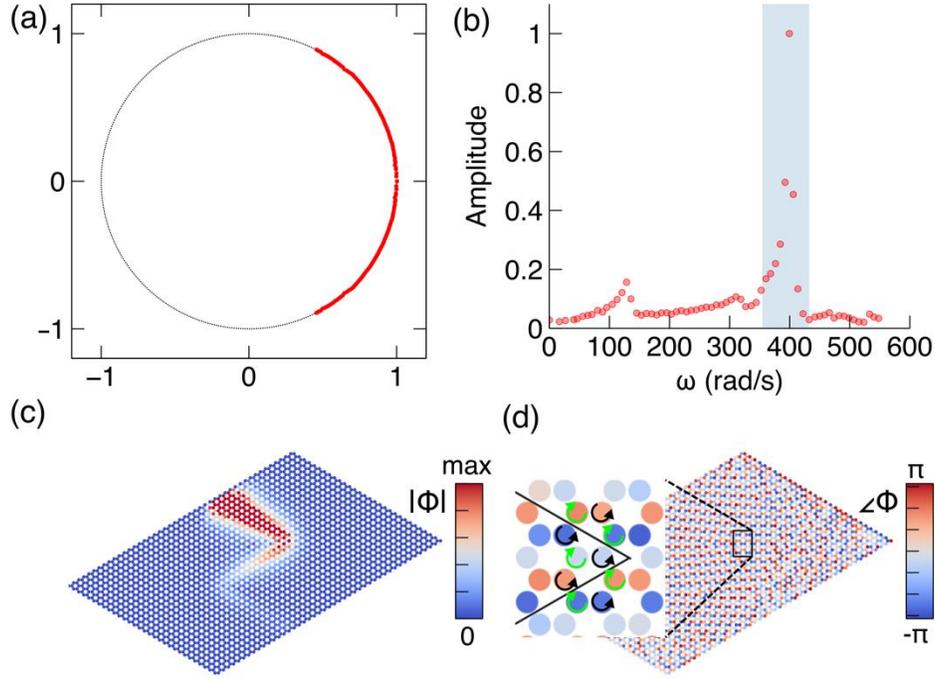

FIG. 2. DMD spectrum and DMD modes. (a) Eigenvalues are visualized on the complex plane located around the unit circle. (b) The mode amplitude varies as a function of frequency. The shading area indicates the bandgap region. (c) The magnitude of the DMD mode with the largest amplitude inside the bandgap region. (d) The phase of the DMD mode with the largest amplitude inside the bandgap region. The inset shows the magnified view around the interface (black line) illustrating the phase difference and valley pseudospin. The black and green arrows show the pseudospin up and pseudospin down indicated by the phase evolution around hexagon corners, respectively.

Using the extracted DMD modes and corresponding time dynamics, we can reconstruct the wave propagation in topological metamaterials using the following expression:

$$\hat{X} = \Phi \Lambda^{t-1} Z \qquad (5)$$

where the diagonal entries of $\Lambda$ contains DMD eigenvalues and $Z = \Phi \backslash x_1$. $x_1$ is the initial condition of our system and backslash is to solve the linear system following the MATLAB notation. Here, we only use the DMD modes inside the bandgap region (10 pairs of DMD modes) and the corresponding eigenvalues to reconstruct the whole process of wave propagation. As displayed in FIG. 3(a), several time-series snapshots represent the wave propagation in our system. Besides, in FIG. 3(b), we also quantify the reconstruction error as a function of duration calculated by $E(t) = \frac{|X(t) - \hat{X}(t)|_2}{|X(t)|_2}$, where $|\cdot|_2$ represents $\ell^2$-norm that is the square root of the sum of the absolute squares of the vector entries. Most relative errors are around 0.97 and oscillating over time. The relative error corresponding to the snapshots shown in FIG. 3(a) is indicated by the arrow in FIG. 3(b). Although the relative error is rather nontrivial, the wave propagation along the Z-shape interface can be clearly observed from the FIG. 3(a) and Supplementary video. It qualitatively captures the evolution dynamics despite the substantially reduced dimensionality of the system, which is visually similar to the original evolution dynamics shown in FIG. 1(b).

DMD can accurately capture the frequency range and characteristics of topological states of elastic topological metamaterials. The nature of the valley pseudospin in our valley system can also be revealed, suggesting that DMD, functioning as a data-driven method, is able to learn the topological nature. Besides, the propagation of topological states can be reconstructed qualitatively only by the DMD modes inside the bandgap and the corresponding time dynamics. Note that DMD with time-delay embedding has demonstrated the ability of increasing the accuracy of reconstruction in several applications [22,23,27–29]. In APPENDIX D and the associated figure, we also show the partial decrease of reconstruction error using the augmented data matrix formed by shift-stacking the original data matrix.

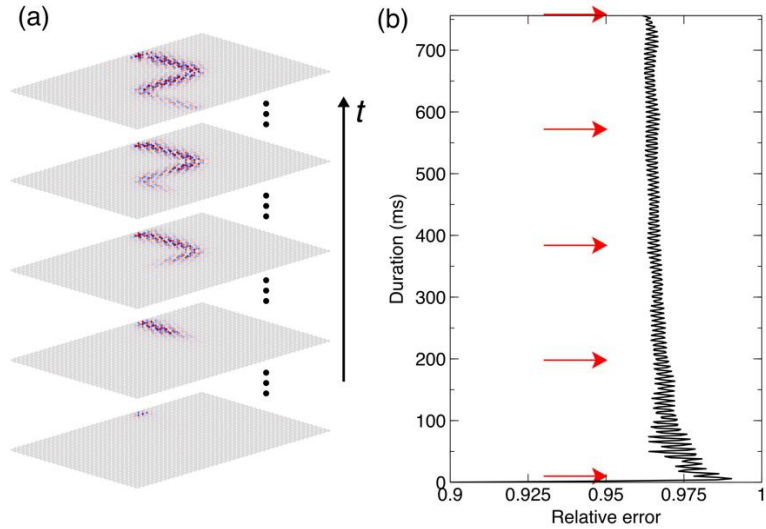

FIG. 3. Reconstruction of propagation of topological states. (a) Snapshots of reconstructed wave propagation represented by horizontal displacement $u_x$ along the Z-shaped interface at $t = 2$ ms, 192 ms, 380 ms, 570 ms and 758 ms. (b) The relative error between the ground truth and reconstruction as a function of duration. The red arrows indicate the relative error of corresponding snapshots in (a).

### III. CLASSIFICATION OF TOPOLOGICAL AND TRADITIONAL METAMATERIALS

Compared with the topological metamaterials, traditional metamaterials function by defect states that cannot support robust transport of elastic waves [30–33]. Here, we demonstrate how to use the extracted DMD modes to classify (and distinguish between) topological and traditional metamaterials. The original DMD modes are high-dimensional and thus difficult to classify using a classification algorithm. Therefore, the feasible way is to find a feature space to project the DMD modes on, resulting in a low-dimensional representation. Specifically, we construct a library of DMD modes inside the bandgap from topological metamaterials and traditional metamaterials $L$:

$$L = \begin{bmatrix} | & | & & | \\ \phi_1 & \phi_2 & \cdots & \phi_N \\ | & | & & | \end{bmatrix} \qquad (6)$$

For the purpose of classification, we consider the absolute value of every element of DMD modes and denote the resulting matrix as $|L|$. To simplify this problem, clusters are determined in one-dimensional principal component space, using the projections of each column of $|L|$ onto the proper principal components of $|L|$:

$$|L| = U_L \Sigma_L V_L^* \qquad (7)$$

With this expression yielding the SVD of the matrix $|L|$ and using

$$P = U_i^T |L| = \Sigma_i V_i^T \qquad (8)$$

where $U_i^T$, $\Sigma_i$ and $V_i^T$ are the transpose of the $i^{th}$ column of $U_L$, $i^{th}$ singular value and the transpose of the $i^{th}$ column of $V_L$, respectively. Note that the transpose of $V_L$ is the same as the conjugate transpose of $V_L$ due to the real value.

The principal components can explain a significant proportion of the variance in the features in topological and traditional metamaterials. Therefore, finding a proper principal component $U_i^T$ is key to distinguish two types of metamaterials. After examining all principal components, we have found that the second principal component is a suitable feature towards the classification task at hand. The second principal component of $|L|$ is shown in FIG. 4(a) and the second principal components of $|L|$ for straight and cross interfaces are shown in FIG. 10(a) and FIG. 10(b). This pattern of principal component shows the features differentiating topological and traditional metamaterials at the beginning of the input port. Therefore, when $|L|$ is projected onto this principal component, the two types of metamaterials can be classified whereas they cannot be classified when $|L|$ is projected onto other principal components. The relevant diagnostic allows us to distinguish topological and traditional metamaterials, corresponding to topologically protected wave propagation and non-topological wave propagation, and hence, accordingly classify them.

As shown in FIG. 4(b), it is obvious that the projected values for topological and traditional metamaterials with different interfaces are separated well and can be simply classified using the *k*-means unsupervised clustering [34]. The classification results and ground truth have 100% agreement. Note that the last DMD mode for Z-shape interface is not an interface state although it is in the bandgap region, so the corresponding projected value does not represent the classification results. Using the same method, the wave propagation along different interfaces (straight and cross interfaces) in topological and traditional metamaterials can also be classified, as detailed in APPENDIX E. The classification results for straight and cross interfaces are demonstrated in the relevant figure therein.

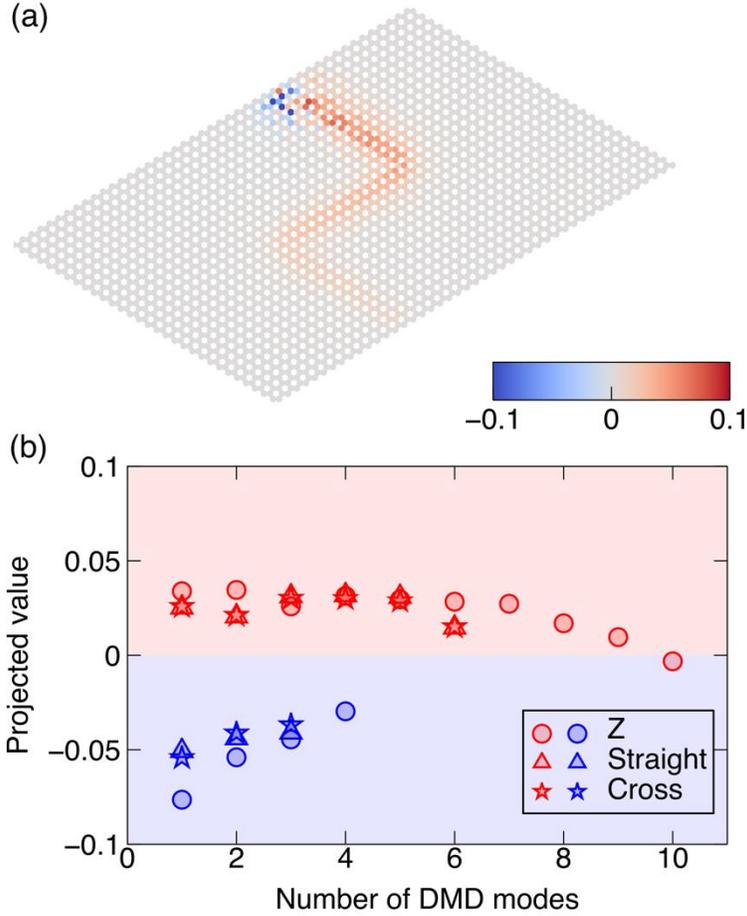

FIG. 4. Classification of topological and traditional metamaterials. (a) The feature space formed by the second principal component of DMD modes with Z-shape interface. (b) The values of each projected DMD mode from different interfaces on the feature space. Red and blue symbols indicate topological and traditional DMD modes, respectively. Circle, triangle and star symbols indicate the Z-shape, straight and cross interfaces.

## IV. PREDICTION OF WAVE PROPAGATION USING DMD

Next, we demonstrate another application of the usefulness of DMD modes in topological metamaterials. Prediction of wave propagation is important when there is lack of data due to the sensor problems or measurement difficulties. Here, we use DMD modes calculated from a part of data (training data) to build a low-dimensional model and further to predict the future propagation of elastic wave given the knowledge of position of interface. Here, as shown in FIG. 5(a), we use the wave propagation in topological metamaterials from 0 ms to 200 ms as the training data. Then, DMD is used to extract the DMD modes inside the bandgap and the corresponding time dynamics (two pairs of DMD modes inside the bandgap are used). Since wave propagation is a process with time and space variation, the DMD modes are limited in space due to the nature of spatial modes, resulting in stoppage of wave and, accordingly, failure of the prediction. A feasible way that we have found relevant towards bypassing this issue is to extend the DMD modes along the given interface and approximate the future wave propagation using the extended DMD modes and the time dynamics from the training data.

First, the least squares method is used in order to identify the time-varying wave velocity $\vec{c}(t)$ by a set of pairs of the positions of wave front and corresponding time. Therefore, the position of wave front can be determined at arbitrary time. Next, after the extracted DMD modes inside the bandgap are reshaped to a matrix form, they are truncated based on the displacement $\vec{d} = \vec{c}(t) \times t$, corresponding to the number of matrix columns, as shown in the training data in FIG. 5(b). The DMD is used again to predict the DMD modes in the future when the elastic wave propagates to the arbitrary position. The prediction time is determined by the length of the given interface. Note that extension by DMD only considers the speed of wave propagation, assuming that it is effectively constant during each segment of the interface, instead of other associated properties such as the dispersive radiation, which is found to be minimal in the present setting. Then, the extended DMD modes $\phi_e$ are shifted according to the shape of the interface, resulting in the shifted DMD mode $\phi_s$. As an example, one of the DMD modes inside the bandgap is shown in FIG. 5(b), where the DMD mode is constrained in space which will cause the stoppage of wave propagation after 200 ms. After being extended by DMD and shifted by the shape of interface, the DMD mode constrained in a certain space can cover the given interface (Z-shaped interface), as shown in FIG. 5(b). Finally, after we extend and shift all DMD modes inside the bandgap from the training data, the time dynamics of the training data are used to predict the wave propagation along the Z-shaped interface, detailed as below:

$$\hat{X} = \Phi_s \Lambda^{t-1} Z_s \tag{9}$$

where $\Phi_s$ is formed by $\phi_s$. The diagonal entries of $\Lambda$ contain DMD eigenvalues from training data and $Z_s = \Phi_s \backslash x_1$. $x_1$ is the initial condition of our system. As shown in FIG. 5(c), several snapshots of predicted wave propagation clearly exhibit the elastic wave traveling along the Z-shape interface. This process is qualitatively similar to the snapshots shown in FIG. 1(b). According to FIG. 5(d), the relative error compared with the reconstructed results in FIG. 1(b) is in the range of 0.95~1, which is acceptable for visualization of future wave propagation.

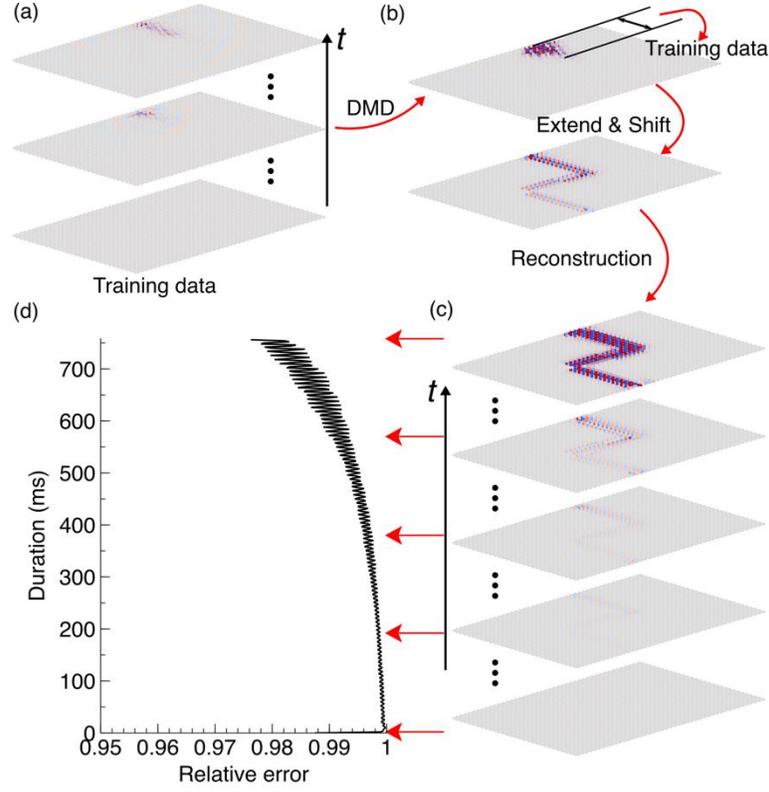

FIG. 5. Prediction of propagation of topological states. (a) The snapshots of horizontal displacement $u_x$ from 0 ms to 200 ms used for training. (b) The top panel displays one of the DMD modes inside the bandgap of the training data. The bottom panel displays the corresponding extended and shifted DMD modes along the Z-shaped interface. (c) The snapshots of the horizontal displacement $u_x$ for the predicted wave propagation at $t$ = 2 ms, 192 ms, 380 ms, 570 ms and 758 ms. (d) The prediction error as a function of duration. The arrows indicate the relative error of the corresponding snapshots in (c).

## V.  CONCLUSIONS AND FUTURE CHALLENGES

In this paper, we provide a guide towards the potential impacts of the application of the DMD method on the wave propagation in topological elastic metamaterials. The analysis of DMD eigenvalues and spectrum show the oscillation and frequency of the DMD modes. Besides, the notable topological interface states and valley pseudospin of the valley system can be reflected by the DMD modes. Furthermore, the reconstruction of topological state propagation is achieved by the low-dimensional model constructed only by the DMD modes inside the bandgap and the corresponding time dynamics. Apart from the fundamental characterization by DMD, we demonstrate the potential that the method bears as concerns the tasks of classification and prediction of wave propagation using DMD modes and associated reduced dynamical descriptions. We put forward a generic feature space to project the DMD modes on for the classification of topological and traditional metamaterials. The prediction of wave propagation along the interface can be achieved by the extension and shift of DMD modes, where the error is at an acceptable level to visualize the future wave propagation. The DMD provides a data-driven method to explore the wave propagation in topological metamaterials and to reveal the potential topological nature. Besides, it opens up an avenue to classify and predict the wave propagation in the pure data-driven approach.

Naturally, this is only a first step along this direction and raises a number of questions that still merit further addressing. One key aspect of interest concerns how to reduce the error. While the results presented herein represent adequate reconstructions (and even predictions) of the time evolution dynamics, it would be highly desirable for such examples to match far more adequately, in a quantitative sense, the real system dynamics. From the point of view of applications, it would be relevant to explore the method in other classes of systems including in ones stemming from higher dimensions and to explore how adequately the method can fare in such more data-intensive settings. Such topics are presently under consideration and associated potential progress will be reported in future publications.

# APPENDIX A: BAND STRUCTURE OF TOPOLOGICAL ELASTIC METAMATERIALS

Our elastic metamaterial is built based on a honeycomb spring-mass system. As displayed in FIG. 6(a), the unit cell is composed of two masses $m_1$ and $m_2$ connected by a spring. The length and the spring constant are $a$ and $k_{spring}$. Therefore, the basic vectors for this unit cell are $\vec{a}_1 = [a_x, -a_y]$, $\vec{a}_2 = [a_x, a_y]$, where $a_x = 3a/2$ and $a_y = \sqrt{3}a/2$. After applying the Bloch's theorem, equations of motion of two masses in one unit cell can be written as:

$$-\omega^2 m_1 \vec{u}^{(1)} = k_{spring}\left[\left(\vec{u}^{(2)} - \vec{u}^{(1)}\right) \cdot \vec{e}_x\right]\vec{e}_x$$
$$+ k_{spring}\left[\left(\vec{u}^{(2)} e^{i\vec{k}\cdot\vec{a}_1} - \vec{u}^{(1)}\right) \cdot \vec{e}_k\right]\vec{e}_k \quad (10)$$
$$+ k_{spring}\left[\left(\vec{u}^{(2)} e^{i\vec{k}\cdot\vec{a}_2} - \vec{u}^{(1)}\right) \cdot \vec{e}_k'\right]\vec{e}_k'$$

$$-\omega^2 m_2 \vec{u}^{(2)} = k_{spring}\left[\left(\vec{u}^{(1)} - \vec{u}^{(2)}\right) \cdot \vec{e}_x\right]\vec{e}_x$$
$$+ k_{spring}\left[\left(\vec{u}^1 e^{-i\vec{k}\cdot\vec{a}_1} - \vec{u}^{(2)}\right) \cdot \vec{e}_k\right]\vec{e}_k \quad (11)$$
$$+ k_{spring}\left[\left(\vec{u}^{(1)} e^{-i\vec{k}\cdot\vec{a}_2} - \vec{u}^{(2)}\right) \cdot \vec{e}_k'\right]\vec{e}_k'$$

where $\vec{e}_x = [1, 0]^T$, $\vec{e}_k = [-\frac{1}{2}, \frac{\sqrt{3}}{2}]^T$ and $\vec{e}_k' = [-\frac{1}{2}, -\frac{\sqrt{3}}{2}]^T$ are three unit vectors along the springs on one mass. The band structure $\omega(k)$ of our system can be obtained by solving the eigenvalue equation in the main text as a function of Bloch wave vector $k$ in the first Brillouin zone. The corresponding eigenmodes $U = \left[u_x^{(1)}, u_y^{(1)}, u_x^{(2)}, u_y^{(2)}\right]$ can also be obtained.

We choose the equal masses on two sites ($m_1 = m_2 = 1$ kg) and $k_{spring} = 10^5$ N/m to find the Dirac point at the corner of the Brillouin zone (K point), as shown in the middle panel of FIG. 6(a). After breaking spatial inversion symmetry by unequal masses on two sites, two bands can be opened to form a bandgap. The left and right panels exhibit the band structure when $m_1 = 0.8$ kg, $m_2 = 1.2$ kg and $m_1 = 1.2$ kg, $m_2 = 0.8$ kg, respectively. At the K valley, eigenmodes corresponding to two bands when $m_1 = 0.8$ kg, $m_2 = 1.2$ kg are $U_1 = \frac{1}{\sqrt{2}}[0,0,1,-i]^T$ and $U_2 = \frac{1}{\sqrt{2}}[1,i,0,0]^T$, while the eigenmodes are $U_1 = \frac{1}{\sqrt{2}}[1,i,0,0]^T$ and $U_2 = \frac{1}{\sqrt{2}}[0,0,1,-i]^T$ after alternating the masses on two sites (see the insets with black arrows that represent the eigenmodes). The obvious band inversion can be seen from the eigen modes of the unit cell.

In FIG. 6(b), the projected band structure is calculated using a sandwiched supercell shown in the right panel of FIG. 6(a). It combines the metamaterials of $m_1 = 0.8$ kg and $m_2 = 1.2$ kg, $m_1 = 1.2$ kg and $m_2 = 0.8$ kg, and $m_1 = 0.8$ kg and $m_2 = 1.2$ kg so that two topological states of different pseudospins corresponding to two types of interfaces emerge inside the bandgap. The excitation frequency of our simulation is 400 rad/s as indicated by the black dashed line in FIG. 6(b).

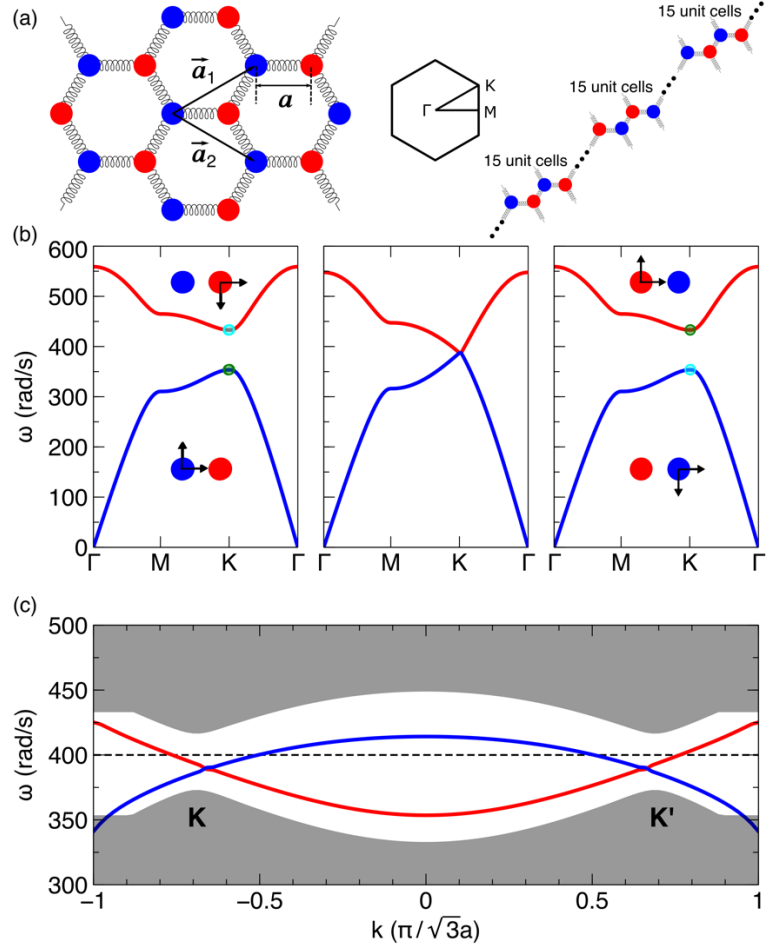

FIG. 6. The band structure of the valley topological metamaterials. (a) The schematic of our elastic metamaterials based on the spring-mass system. The basic vectors of unit cell are shown in $\vec{a}_1$ and $\vec{a}_2$. The length of the spring is $a$. The first Brillouin zone with high symmetry points Γ, M and K is shown in the middle panel. The sandwiched supercell for the calculation of projected band structure is shown in the right panel. (b) From left to right, the band structure when $m_1 = 0.8$ kg, $m_2 = 1.2$ kg, $m_1 = 1$ kg, $m_2 = 1$ kg and $m_1 = 1.2$ kg, $m_2 = 0.8$ kg are shown. The eigenmodes corresponding to the K valley are shown in the first panel and the third panel. Motions along horizontal and vertical directions are marked on the two sites. (c) The projected band structure with two topological states inside the bandgap. The black dashed line indicates the excitation frequency of the simulation setup.

**APPENDIX B: THE TRUNCATION OF SINGULAR VALUES OF THE DATA MATRIX**

Choosing the proper truncation of the singular value decomposition of $X$ is important to obtain the best-fit linear operator $A$. To identify the truncation $r$ of the singular values, we wish to ensure the minimization of the reconstruction error. The reconstruction $\hat{X}$ is conducted by DMD modes inside the bandgap region and corresponding time dynamics and is compared with the original wave propagation. In FIG. 7(a), the map of the reconstruction error calculated by $E(t) = \frac{|X(t)-\hat{X}(t)|_2}{|X(t)|_2}$ is given as a function of the number of singular values (truncation $r$) and duration. The singular value spectrum shows that singular values decay slowly, indicating that many modes are needed. Accordingly, with the increase of the number of singular values in a certain range (1~140), the reconstruction error does not change significantly in the logarithm scale. However, when the number of singular values further increases, the reconstruction error will significantly increase.

Therefore, the proper truncation is in the range of 1 to 140, which we magnify in the FIG. 7(b). As the number of singular values increases, the reconstruction error will decrease to a minimum. We choose the $r = 131$ as the number of singular values corresponding to the inflection point in the singular value spectrum.

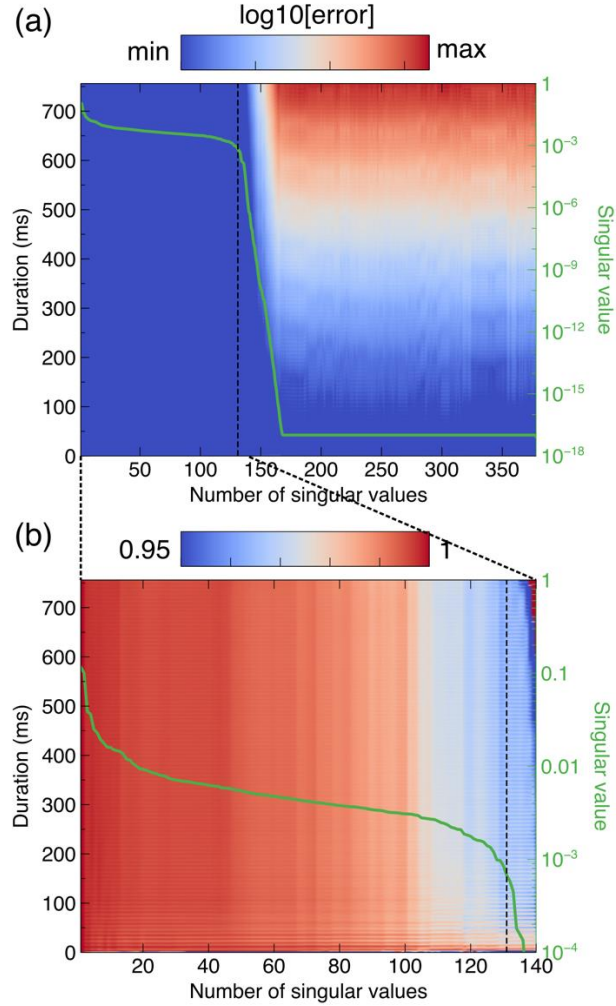

Fig. 7. (a) The reconstruction error as a function of truncation and duration. The error is shown on a logarithmic scale. (b) The zoom-in reconstruction error as a function of truncation from 1 to 140 and duration. The vertical dashed lines in (a) and (b) correspond to the selected $r = 131$.

**APPENDIX C: DMD SPECTRA AND MODES OF WAVE PROPAGATION ALONG INTERFACE WITH DIFFERENT SHAPES**

FIG. 8(a) gives the DMD spectrum indicating the relation between the frequency and the mode amplitude for the topological states propagation along the straight interface (configuration shown in the inset). Similar to FIG. 2(b), it is clear that the region with high mode amplitude corresponds to the bandgap region (shaded area). The mode inside the bandgap region with the largest amplitude is chosen as the prototypical mode of interest (and of relevance to the dynamics). FIG. 8(b) exhibits the magnitude and phase of this dynamic spatial mode of our system. The interface state can be observed from the magnitude of the DMD modes. The displacement is concentrated along the straight interface and decays rapidly away from the interface. Besides, the elastic wave can travel along the interface with bends. The phases of the DMD modes also reflect the characteristics of topological states. The distribution of phase along the interface has a certain pattern, representing the valley pseudospin of our system as described in the main text. Likewise, in FIG. 8(c) and FIG. 8(d), we calculate the DMD spectrum and DMD modes inside the bandgap region with the largest amplitude for the topological states propagation along the cross interface (configuration shown in the inset). The DMD mode in FIG. 8(d) shows that the elastic wave travels along the path at the beginning and when it arrives at the intersection, it propagates to two sides instead of the straight path. Because of the valley-locking effect, the wave will propagate along certain interface with same valley projection [11,35–37]. The generated elastic wave is projected by the K valley according to the group velocity in projected band structure (FIG. 6(c)). Therefore, the elastic wave will only propagate along the K-valley-projected topological interfaces. Note that apart from DMD modes shown in FIG. 8(b) and FIG. 8(d) which have the largest amplitudes in the DMD spectra, other DMD modes inside the bandgap region are also interface states.

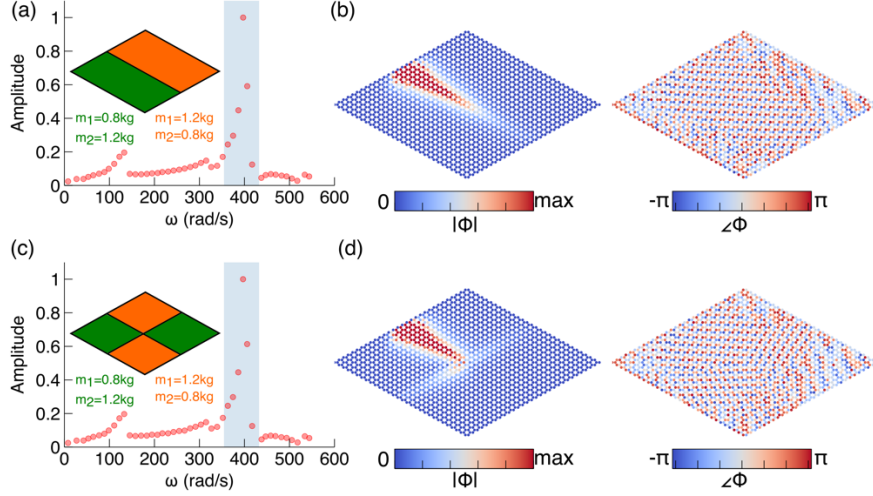

Fig. 8. DMD spectrum and DMD modes of straight interface and cross interface are shown in (a), (b) and (c), (d), respectively. The insets in (a) and (c) are the configurations of elastic topological metamaterials.

## APPENDIX D: DYNAMIC MODE DECOMPOSITION WITH TIME-DELAY EMBEDDING

Recently, the approach of time-delay embedding has been shown to be a general method to generate proper observable measurements to render the reconstruction more accurate as discussed in the main text. By embedding future temporally consecutive snapshots into the current snapshot, time-delay embedding augments the limited spatial observables and provides extra observables. The DMD with time-delaying embedding can be achieved by the augmented data matrix $X_{aug}$ by shift-stacking the original data matrix as shown below:

$$X_{aug} = \begin{bmatrix} | & | & & | \\ x_1 & x_2 & \cdots & x_{m-h} \\ | & | & & | \\ | & | & & | \\ x_2 & x_3 & \cdots & x_{m-h+1} \\ | & | & & | \\ | & | & & | \\ x_h & x_{h+1} & \cdots & x_{m-1} \\ | & | & & | \end{bmatrix} \qquad (12)$$

where $h$ is the number of stacks. $X'_{aug}$ can be induced likewise. Using the augmented data matrix to conduct DMD, the reconstruction error can be reduced. As shown in FIG. 9, the relative error calculated by $E = \frac{|X-\hat{X}|_F}{|X|_F}$ decreases with the increase of the number of stacks and becomes saturated at around 0.65 even with a larger number of stacks. However, when the number of stacks increases, the augmented data matrix becomes large, leading to a heavy computation cost. Therefore, there is a tradeoff balance between accuracy and efficiency in the case of real-world applications and practitioners should seek to strike a relevant balance to that effect.

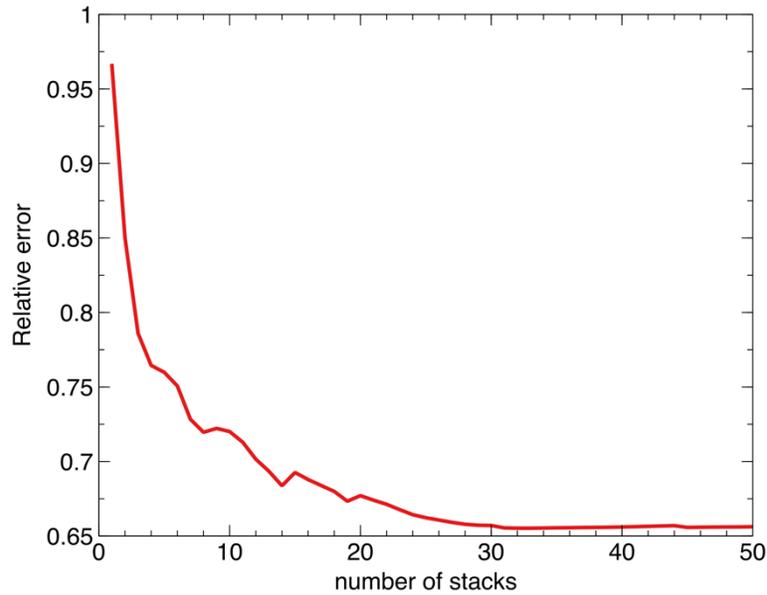

FIG. 9. The relative error of reconstruction as a function of number of stacks.

# APPENDIX E: CLASSIFICATION OF TOPOLOGICAL AND TRADITIONAL METAMATERIALS WITH DIFFERENT INTERFACES

For classification of topological and traditional metamaterials with different interfaces, we follow the generic method we introduce in the main text. The principal components can explain a significant proportion of the variance in the features in topological and traditional metamaterials. Therefore, we extract the principal components of the library composed of DMD modes of topological and traditional metamaterials with different interfaces. As for different shapes of the interfaces, the feature space from the principal components for the DMD modes to project on can be determined as visualized in FIG. 10(a) and FIG. 10(c), corresponding to the straight interface and cross interface. Similarly, the feature spaces are also the second principal components in the singular value decomposition, in which the main difference between topological wave propagation and non-topological wave propagation is reflected, and hence can distinguish two types of metamaterials. After the DMD modes inside the bandgap for both topological metamaterials and traditional metamaterials are projected onto the feature space, the DMD modes lead to a scalar value. As shown in FIG. 10(b) and FIG. 10(d), projected values of topological and traditional DMD modes are separated in the feature space and can be classified using the $k$-means unsupervised clustering. For both topological propagation along straight interface and cross interface, the classification results and ground truth have 100% agreement.

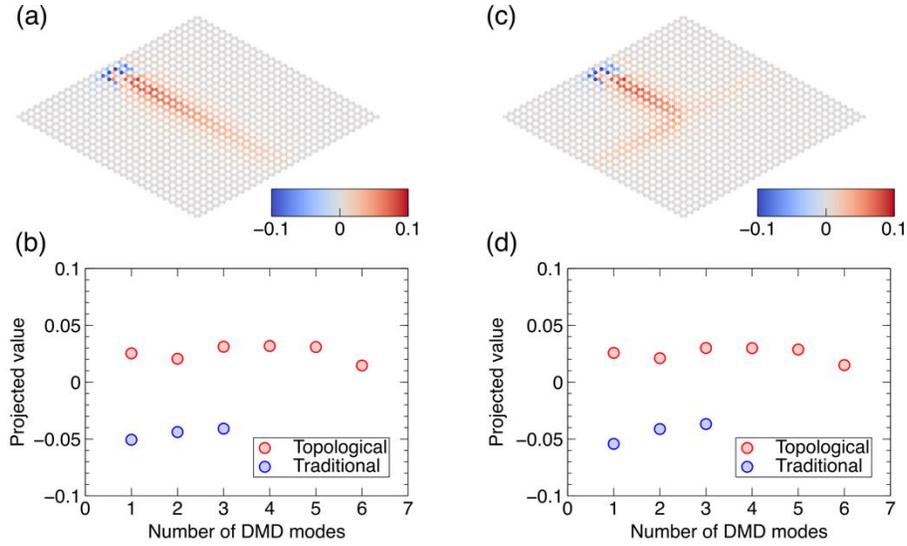

Fig. 10. Classification of topological and traditional waveguides with different interface. (a) (c) The synthetic feature space of topological metamaterials with straight interface and cross interface. (b) (d) The values of each projected DMD mode on the feature space, corresponding to (a) and (c). Red and blue circles indicate topological and traditional DMD modes, respectively.

**ACKNOWLEDGEMENTS**

The present paper is based on work that was supported by the US National Science Foundation under Grant No. EFRI-1741685 (SL and JY), DMS-1809074 (PGK), and DMS-2204702 (PGK).